\documentclass[aps,twocolumn,nofootinbib]{revtex4}
\usepackage[latin1]{inputenc}
\usepackage{epsfig}
\newcommand{\beq}{\begin{equation}}
\newcommand{\eeq}{\end{equation}}

\begin{document}

\title{Quantum Fokker-Planck structure of the Lindblad
equation}

\author{Mário J. de Oliveira}
\affiliation{Universidade de São Paulo, Instituto de Física,
Rua do Matão, 1371, 05508-090 São Paulo, SP, Brazil}

\begin{abstract}

We show that the quantum Fokker-Planck equation,
obtained by a canonical quantization of its classical
version, can be transformed into an equation of the 
Lindblad form. This result allows us to conclude that
the quantum Fokker-Planck equation preserves
the trace and positivity of the density operator.
The Fokker-Planck structure gives explicit
expression for the quantum equivalence of probability
current as well as the quantum equivalence of
detailed balance. We also propose expression for the
rate of entropy production and show that it does
not vanish for a closed system except in equilibrium.
 
\end{abstract}

\maketitle

\section{Introduction}

The dynamics of quantum open systems 
\cite{lindblad1976,iche1978,dekker1979a,dekker1979b,%
dekker1981,caldeira1981,caldeira1983a,caldeira1983b,%
chang1985,gallis1993,munro1996,gardiner1999,banik2002,%
breuer2002,caldeira2014,manzano2020}
is usually formulated considering a system and its
environment. The equations of motion of the system
is then derived by summing out the degrees of freedom
of the environment. The derivation is not accomplished
without assuming an approximation concerning the
interactions of the system with its environment.
Usually, the environment is considered a thermal system
which means to say that the interaction with the system
is regarded as being of stochastic nature. 
Starting from the quantum Liouville equation for the total
system, which is the system proper and its environment,
the resulting evolution equation is a quantum Liouville
equation supplemented by a dissipation term $D$
\cite{breuer2002,caldeira2014,manzano2020},
\beq
\frac{d\rho}{dt} = \frac1{i\hbar}[H,\rho] + D,
\label{5}
\eeq
where $H$ is the Hamiltonian of the system.

Taking into account that the variables of the
environment acts as stochastic variables, the
reduced equation (\ref{5}) describes a quantum
Markov process which means that the dynamics
is described by a quantum dynamic semi-group.
The most general form of the generator $D$ that
has the property of semi-group and that preserves
the trace and positivity of the density operator
is of the Lindblad form
\cite{breuer2002,caldeira2014,manzano2020},
\beq
D = \sum_{jk} a_{jk} (2 A_j\rho A_k^\dagger
- A_k^\dagger A_j\rho - \rho A_k^\dagger A_j),
\label{6}
\eeq
where $a_{jk}$ are the entries of a Hermitian
and positive matrix. The equation (\ref{5}) with
$D$ in the form (\ref{6}) is called the Lindblad
equation.

An alternative approach to reach an equation for
quantum open system is to consider the classical
Fokker-Planck-Kramers equation
\cite{kampen1981,risken1989,gardiner2009,tome2015},
which is
known to describe open classical systems, and
carry out its canonical quantization
\cite{oliveira2016,oliveira2018,tome2023}. 
The resulting quantum Fokker-Planck (FP) equation
that gives the time evolution of the density
operator $\rho$ is given by
\beq
i\hbar \frac{d\rho}{dt} = [H,\rho]
- \frac12\sum_j [x_j, J_j + J_j^\dagger],
\label{9}
\eeq
where $J_j$ is the quantum version of the probability
current, given by 
\beq
J_j = -\gamma_j(\rho g_j 
+\frac{m}{i\hbar\beta_j} [x_j,\rho]),
\eeq
and
\beq
g_j = -\frac{m}{i\hbar \beta_j}
(e^{\beta_j H}x_je^{-\beta_j H} -x_j).
\label{20}
\eeq
As happens to its classical version, this equation
describes the contact of a system of interacting
particles of mass $m$ with thermal reservoirs at
temperatures inversely proportional to $\beta_j$, and
$\gamma_j$ measures the strength of the interaction
with the reservoirs. The positions and momenta of the
particles are denoted by $x_j$ and $p_j$. The first
and second terms of the current $J_j$ corresponds to
the dissipation and fluctuation, respectively. We
remark that $g_j$, which is related to dissipation,
is not in general proportional to the momentum $p_j$
as in the classical dissipation but becomes
proportional to $p_j$ in the classical limit.

Equations similar to (\ref{9}) were considered by
Dekker \cite{dekker1981} and by Caldeira and 
Leggett \cite{caldeira1983b,caldeira2014}. However,
there is a difference in that the dissipation term
of their equations is proportional to the momentum 
whereas in equation (\ref{9}), the dissipation 
term is proportional to a general term $g_j$ which
depends on the Hamiltonian of the system as can be
seen in equation (\ref{20}). This form of $g_j$ is
crucial if we wish to describe the thermodynamic
equilibrium, or in other words, if we wish that
the system thermalizes in the long run.

The quantum FP equation (\ref{9}) 
can be written in a more symmetric form
in terms of annihilation and
creation operator $a_j$ and $a_j^\dagger$ in 
which case it reads 
\beq
i\hbar \frac{d\rho}{dt} = [H,\rho]
- \sum_j ([a_j,J_j^\dagger] + [a_j^\dagger,J_j]),
\label{10}
\eeq
where 
\beq
J_j = i\gamma_j(g_j\rho + \frac1{\beta_j}[a_j,\rho]),
\eeq
and
\beq
g_j = \frac1{\beta_j}(e^{-\beta H}a_j e^{\beta_j h} - a_j).
\eeq
It is worth writing $J_j$ in the form
\beq
J_j = \frac{i\gamma_j}{\beta_j}
(e^{-\beta H}a_j e^{\beta_j h}\rho -\rho a_j).
\label{18}
\eeq

The quantum FP equation in either forms (\ref{9})
or (\ref{10}) is understood as describing a quantum
Markov process and in this sense it should be of
the type given by equations (\ref{5}) and (\ref{6}).
The first term of the quantum FP equation which
corresponds to a unitary transformation is indeed
the same. As to the second non-unitary term, it
is not of the Lindblad type given by (\ref{6}).
The main purpose of the present paper is to show
that the second term of the quantum FP equation
can be transformed into the Lindblad form, showing
thus that the quantum FP equation preserves the
trace and positivity of the density operator $\rho$.

\section{Quantum FP structure}

We consider a Hilbert vector space and choose a
basis consisting of the eigenvectors of some
Hermitian operator $L$. The eigenvectors
associated to the eigenvalue $\lambda_i$ of 
$L$ are denoted by $\phi_i$, that is,
$L\phi_i = \lambda_i\phi_i$.
Taking into account that $L$ is Hermitian
its left eigenvectors are the adjoint vector 
$\phi_i^\dagger$.
The operators acting on the vectors of the
Hilbert space such as $L$  
itself can also be understood as belonging
to another vector space, the Liouville space,
whose complete basis consists of the operators
$A_{ij}=\phi_j \phi_i^\dagger$.

The general expression of a non-unitary generator 
$D$ which  preserve the trace of $\rho$ and its
complete positivity for any initial condition
is of the Lindblad form \cite{breuer2002}, 
\beq
D = \sum_{ij,k\ell} a_{ij,k\ell}
(2 A_{ij}\rho A_{k\ell}^\dagger
- A_{k\ell}^\dagger A_{ij}\rho
- \rho A_{k\ell}^\dagger A_{ij}),
\eeq
where $a_{ij,k\ell}$ are the entries of a Hermitian
and positive matrix. Our point of depart is the
following expression of the Lindblad type
\[
D = \sum_{ij,k\ell} b_{ij,k\ell}
(2 A_{ij}\rho A_{k\ell}^\dagger
- A_{k\ell}^\dagger A_{ij}\rho
- \rho A_{k\ell}^\dagger A_{ij})
\]
\beq
+ \sum_{ij,k\ell} c_{ij,k\ell}
(2 A_{ij}^\dagger\rho A_{k\ell}
- A_{k\ell} A_{ij}^\dagger\rho
- \rho A_{k\ell} A_{ij}^\dagger),
\eeq
where $b_{ij,k\ell}$ and $c_{ij,k\ell}$ are the
entries of Hermitian and positive matrices, and
are nonzero only when $i\leq j$ and $k\leq \ell$. 

Defining the operators $B_{ij}$ by
\beq
\alpha_{ij}B_{ij}
= \sum_{k\ell} b_{ij,k\ell}^* A_{k\ell},
\label{16b}
\eeq
where $\alpha_{ij}\geq0$,
the first summation can be written in the form
\beq
\sum_{ij}\alpha_{ij}(A_{ij}\rho B_{ij}^\dagger
+ B_{ij}\rho A_{ij}^\dagger
- A_{ij}^\dagger B_{ij}\rho - \rho B_{ij}^\dagger A_{ij}).
\eeq
In an analogous manner we defined the operators $C_{ij}$ by
\beq
\alpha_{ij}C_{ij} = \sum_{k\ell}c_{ij,k\ell} A_{k\ell},
\label{16c}
\eeq
and the second summation becomes
\beq
\sum_{ij}\alpha_{ij}(A_{ij}^\dagger \rho C_{ij}
+  C_{ij}^\dagger\rho A_{ij} - A_{ij} C_{ij}^\dagger\rho 
- \rho C_{ij} A_{ij}^\dagger).
\eeq
Summing up these two terms, we reach the expression
\beq
D = \sum_{ij}\alpha_{ij}\{
[A_{ij}, \rho  B_{ij}^\dagger - C_{ij}^\dagger\rho] 
- [A_{ij}^\dagger,B_{ij}\rho - \rho C_{ij}]\},
\eeq
and the Lindblad equation acquires the FP
structure
\beq
i\hbar \frac{d\rho}{dt} = [H,\rho] - \sum_{ij}\{
[A_{ij}, J_{ij}^\dagger] + [A_{ij}^\dagger,J_{ij}]\},
\label{15}
\eeq
where $J_{ij}$ is given by
\beq
J_{ij} = i\hbar \alpha_{ij}(B_{ij}\rho - \rho C_{ij}).
\label{17}
\eeq

The expression (\ref{15}) for the Lindblad equation is
particularly meaningful because $J_{ij}$ represents
the quantum version of the probability current.
Suppose that the right-hand side
of (\ref{15}) vanishes for a density operator $\rho_0$,
in which case the system is said to be in a stationary
state. If in addition the currents $J_{ij}(\rho_0)$ vanish,
in which case $[H,\rho]$ also vanishes, then the
system will be in thermodynamic equilibrium.
The vanishing of the currents corresponds to
the condition of detailed balance since each
term of the summation in (\ref{15}) vanishes.
The condition of detailed balance is represented by
\beq
B_{ij}\rho_0 = \rho_0  C_{ij},
\label{19}
\eeq
for some $\rho_0$.

\section{Contact with thermal reservoirs}

There are various possibilities of choosing the
operators $B_{ij}$ and $C_{ij}$. The only restriction
is that the coefficients of the expansions (\ref{16b})
and (\ref{16c}) define Hermitian and positive matrices.
The choices will depend on the type of physical
conditions one wants to describe. Here we choose
these operators with the purpose of describing
a system in contact with several heat reservoir at
distinct temperatures. When the temperatures are
all the same, then in the stationary state the
system will be in thermodynamic equilibrium in
which case the density operator is of the 
Gibbs form
\beq
\rho_0 = \frac1Z e^{-\beta H},
\eeq
where $\beta$ is inversely proportional to the
temperature of the reservoirs.

We choose $C_{ij}$ and $B_{ij}$ so that 
\beq
B_{ij}e^{-\beta_{ij}H} = e^{-\beta_{ij}H}C_{ij},
\label{41}
\eeq
where $\beta_{ij}$ are constants and $H$ is
the Hamiltonian. When $\beta_{ij}$ has the same
value, independent of $i$ and $j$,
the condition (\ref{41}) guarantees that 
the detailed balance condition (\ref{19}) is
fulfilled and the system will be found
in thermodynamic equilibrium. In other words, the
system will thermalize  in the long run.

A simplification arises by choosing
$C_{ij}=A_{ij}$ so that $B_{ij}$ is given by
\beq
B_{ij} = e^{-\beta_{ij}H}A_{ij}e^{\beta_{ij}H},
\label{29}
\eeq
and the current (\ref{17}) becomes
\beq
J_{ij} = i\hbar \alpha_{ij} (e^{-\beta_{ij}H}A_{ij}
e^{\beta_{ij}H}\rho - \rho  A_{ij}),
\eeq
which has the form (\ref{18}) as desired.

It is now left to show that the coefficients of the
expansion of $B_{ij}$ in terms of $A_{ij}$ 
are entries of a Hermitian and positive matrix.
We recall that $A_{ij}=\phi_j \phi_i^\dagger$ 
where $\phi_i$ are the eigenvectors of $L$. 

Let us denote by $\chi_j$ and $E_j$ the eigenvectors
and eigenvalues of the Hamiltonian, that is,
$H\chi_i= E_i\chi_i$. The operators
$X_{ij}=\chi_j \chi_i^\dagger$ can be considered
a complete basis of the Liouville space.
The change from this basis to the basis
used above is given by the unitary transformation
\beq
A_{ij} = \sum_{k\ell} U_{ij,k\ell} X_{k\ell}.
\eeq
Replacing this expression in (\ref{29}), 
we find
\beq
B_{ij} = \sum_{ij} U_{ij,k\ell}
e^{-\beta_{ij}(E_k-E_\ell)} X_{k\ell}.
\eeq
Using 
\beq
X_{ij} = \sum_{k\ell} U_{ij,k\ell}^\dagger A_{k\ell},
\eeq
it can be written as
\beq
B_{ij} = \sum_{k\ell} G_{ij,k\ell}A_{k\ell},
\eeq
where
\beq
G_{k\ell,ij}^*  =  \sum_{mn}
U_{ij,mn} e^{-\beta_{ij}(E_m-E_n)} U_{mn,k\ell}^\dagger.
\eeq
From this expression it follows that 
$G_{k\ell,ij}^* =G_{ij,k\ell}$ and that
the matrix with elements $G_{ij,k\ell}$
is positive, finishing our demonstration.

\section{Entropy production}

The Lindblad equation or its quantum FP version
(\ref{15}) is supposed to describe the thermodynamic
of quantum system in equilibrium or out of equilibrium.
In this sense it constitutes the basic equation of
a stochastic quantum thermodynamics
\cite{oliveira2016,oliveira2018,tome2023}.
A fundamental concept in the description of systems
out of thermodynamic equilibrium is the entropy
production which we discuss below.

The average energy $U$ is defined by
\beq
U={\rm Tr}(H\rho),
\eeq
and its time evolution is obtained from the quantum FP
equation (\ref{15}), and is given by
\beq
\frac{dU}{dt} = \sum_{ij} \Phi_{ij}^u,
\eeq
where
\beq
\Phi_{ij}^u= - \frac1{i\hbar }\{
{\rm Tr}[H,A_{ij}] J_{ij}^\dagger
+ {\rm Tr}[H,A_{ij}^\dagger]J_{ij}\},
\eeq
and $\Phi^u $ is understood as the flux of energy
to the system.
The entropy of the system is defined by
\beq
S = -k{\rm Tr}(\rho\ln\rho),
\eeq
and its time variation is obtained from the FP equation
(\ref{15}), and is given by
\beq
\frac{dS}{dt} =\Pi + \Phi,
\eeq
where
\beq
\Pi= \frac{k}{i\hbar} \sum_{ij}{\rm Tr}(
[\ln\rho-\ln\rho_{ij},A_{ij}] J_{ij}^\dagger
+ [\ln\rho-\ln\rho_{ij},A_{ij}^\dagger]J_{ij}),
\eeq
is understood as the rate of entropy production, and 
\beq
\Phi= \frac{k}{i\hbar} \sum_{ij}{\rm Tr}(
[\ln\rho_{ij},A_{ij}] J_{ij}^\dagger
+ [\ln\rho_{ij},A_{ij}^\dagger]J_{ij})
\eeq
is understood as the flux of entropy to the system,
where $\rho_{ij}$ is given by
\beq
B_{ij}\rho_{ij} = \rho_{ij}C_{ij}.
\eeq

For the case of a system in contact with several 
heat reservoirs, $\ln\rho_{ij}$ is proportional 
to $-\beta_{ij}H$ and the flux of entropy can be
written in the form
\beq
\Phi = \sum_{ij}\frac1{T_{ij}} \Phi_{ij}^u,
\eeq
where $T_{ij}=1/k\beta_{ij}$ and can be understood as
the temperature of the heat reservoirs. We recall
that $\Phi_{ij}^u$ is the energy flux, or heat flux
in the present case, from each heat reservoirs to
the system

In the stationary state, the total flux of energy
\beq
\Phi^u = \sum_{ij} \Phi_{ij}^u
\eeq
vanishes, but it does not mean that each flux
$\Phi_{ij}^u$ vanishes because the temperatures
are not all the same. In this case the flux of entropy
$\Phi$ does not vanish, and $\Pi=\Phi$. As a consequence, 
in the nonequilibrium stationary state the production
of entropy $\Pi$ is nonzero. 
If however temperatures of the reservoirs are all the
same, $T_{ij}=T$, then
\beq
\Phi = \frac1{T} \Phi^u,
\eeq
and in this case $\Phi$ vanishes and so does $\Pi$,
which describes a system in thermodynamic equilibrium.

\section{Isolated system}

Usually one describes an isolated system by
the Liouville equation. This is in fact the
point of depart of deriving the Lindblad equations
for a given system. The given system plus the
environment are assumed to be described by the
Liouville equation because as a whole they are isolated,
and the total energy is a conserved quantity.
If the system of interest is itself
isolated, then we could describe it by the
Liouville equation, which means to regard the
dissipative term $D$ of equation (\ref{5})
as nonexistent.

The mean feature of an isolated system that allows
us to use the Liouville equation is that the
Hamiltonian is strictly conserved along a 
trajectory. However, it is possible to impose
a conservation of the Hamiltonian along 
a stochastic trajectory in such a way that the dissipation
term does not need to be absent. Indeed, if
we choose the operators $C_{ij}=B_{ij}=A_{ij}$ then 
\beq
J_{ij}(\rho) = i\hbar \alpha_{ij}[A_{ij},\rho],
\eeq
and if $A_{ij}$ commutes with the Hamiltonian
$J_{ij}(H)=0$ and the Hamiltonian will be
strictly invariant.

In this case, $\Phi_{ij}^u$ vanishes identically
and there will be no flux of energy, as expected.
The fluxes of entropy $\Phi_{ij}$ will also 
vanish identically and there is no flux of
entropy to or from the system.
The rate of entropy production $\Pi$ equals 
$dS/dt$ and is given by
\beq
\Pi= \frac{k}{i\hbar} \sum_{ij}{\rm Tr}(
[\ln\rho,A_{ij}] J_{ij}^\dagger
+ [\ln\rho,A_{ij}^\dagger]J_{ij}),
\eeq
It is nonzero but vanishes in the stationary state
in which case it is also the equilibrium state
because $J_{ij}$ vanishes.

\section{Conclusion}

We have shown that the quantum FP equation 
can be transformed into an equation that 
has the Lindblad equation. As the Lindblad
equations preserves the trace and positivity
of the density operator so does the quantum
FP equation. The advantage of the FP form
is that one easily recognizes the quantum equivalents
of the probability current and of the
detailed balance condition. When the detailed
balance condition is not satisfied, the quantum
system in the long run will be found in a nonequilibrium
stationary state. In this case the production of
entropy is nonzero and can be obtained by the
expression provided for the rate of entropy 
production. The Fokker-Planck form allows to
determine a dissipation term for the case
in which the Hamiltonian of the system is
strictly constant, which can be understood as
a closed system.


\end{document}